\documentclass[conference]{IEEEtran}

\newif\iffull
\fulltrue

\usepackage{algorithmic, algorithm}
\usepackage[english]{babel}
\usepackage{amssymb}
\usepackage{amsmath}
\usepackage{xcolor}
\usepackage{cite}
\usepackage{graphicx}
\usepackage{amsthm}
\usepackage[hidelinks]{hyperref}

\theoremstyle{definition}
\newtheorem{theorem}{Theorem}
\newtheorem{lemma}{Lemma}
\newtheorem{definition}{Definition}
\newtheorem{corollary}{Corollary}
\newtheorem{problem}{Problem}
\newtheorem{construction}{Construction}
\newtheorem{example}{Example}

\newcommand{\set}[2]{\left\{#1\;\left|\; #2\right.\right\}}
\newcommand{\abs}[1]{\left|#1\right|}
\renewcommand{\v}[1]{\mathbf{#1}}

\newcommand{\full}[1]{\iffull #1 \fi}
\newcommand{\short}[1]{\iffull \else #1 \fi}

\title{Codes for Constrained Periodicity}

\author{\IEEEauthorblockN{Adir Kobovich\IEEEauthorrefmark{1}, Orian Leitersdorf\IEEEauthorrefmark{1}, Daniella Bar-Lev, and Eitan Yaakobi}
\IEEEauthorblockA{\textit{Technion -- Israel Institute of Technology, Haifa, Israel} \\
Email: \{adir.k, orianl\}@campus.technion.ac.il, \{daniellalev, yaakobi\}@cs.technion.ac.il}
\IEEEauthorblockA{\IEEEauthorrefmark{1}These authors contributed equally to this work.}}

\IEEEoverridecommandlockouts
\IEEEaftertitletext{\vspace{-5pt}}

\begin{document}

\maketitle

% ---- Abstract ---- %
\begin{abstract}
Reliability is an inherent challenge for the emerging nonvolatile technology of racetrack memories, and there exists a fundamental relationship between codes designed for racetrack memories and codes with \emph{constrained periodicity}. Previous works have sought to construct codes that avoid periodicity in windows, yet have either only provided existence proofs or required high redundancy. This paper provides the first constructions for avoiding periodicity that are both efficient (average-linear time) and with low redundancy (near the lower bound). The proposed algorithms are based on \emph{iteratively repairing} windows which contain periodicity until all the windows are valid. Intuitively, such algorithms should \emph{not} converge as there is no monotonic progression; yet, we prove convergence with average-linear time complexity by exploiting subtle properties of the encoder. Overall, we both provide constructions that avoid periodicity in all windows, and we also study the cardinality of such constraints.
\end{abstract}

% ---- Introduction ---- %
\section{Introduction}
\label{sec:introduction}

Racetrack memories are an emerging form of nonvolatile memory with extremely high density that have the potential to overcome the fundamental constraints of traditional memory devices~\cite{Racetrack2008, RacetrackPIEEE}. They consist of magnetic nanowires that store numerous bits through magnetic polarity; their value is accessed by shifting the bits stored in each wire to heads at fixed locations. Unfortunately, the shift operation is highly unreliable, thereby leading to position errors in the form of deletions and sticky insertions~\cite{HiFi}. Codes that address these errors have a fundamental relationship to \emph{constrained periodicity} due to the reading structure involving multiple heads simultaneously reading at fixed offsets~\cite{chee2018coding, CheeReconstruction}.

This paper aims to develop efficient codes that constrain periodicity in all windows of encoded messages. Specifically, we consider both the \emph{$\ell$-window $p$-period avoiding} (\emph{PA}) \emph{constraint} where all windows of length $\ell$ cannot contain a period $p$, and the \emph{$\ell$-window $p$-least-period avoiding} (\emph{LPA}) \emph{constraint} where all windows of length $\ell$ cannot contain any period $p' < p$. These constraints were first considered by Chee~\textit{et al.}~\cite{chee2018coding}, where a lower bound on the cardinality proved the \emph{existence} of a binary LPA code with $\ell = \lceil \log(n) \rceil + p + 1$ using a \emph{single} redundancy symbol. Sima and Bruck~\cite{MultipleHeadRacetrack} later proposed an $O(n^2 p \log n)$ time algorithm for the constraint $\ell = \lceil \log(n) \rceil + 3p - 2$ using $p + 1$ redundancy symbols; yet, there remains a significant gap in the redundancy between this explicit construction and the lower bound provided by Chee~\textit{et al.}~\cite{chee2018coding}. Conversely, in this paper, we propose an $O(n)$ average-time construction of LPA codes using a single redundancy symbol for $\ell$ being the minimal integer satisfying $\ell = \lceil \log (n-\ell+2) \rceil + p + 1$. Further, we prove that LPA codes that use a single redundancy symbol exist only for values of $\ell$ that satisfy $\ell \geq \log(n-2\ell+p) + p - 3.5$.

The proposed approach is based on \emph{iteratively repairing} invalid windows until a legal message is encountered. Specifically, as long as there exists a window with invalid periodicity, we remove the window and append an alternate encoding of the window (of identical length). While intuitively this algorithm should not converge due to the lack of monotonic progression (e.g., appended symbols may create additional periodic windows) and the existence of cycles (e.g., repairing an invalid window may lead to the original message), we show that subtle properties of the algorithm guarantee convergence. Further, we prove that only $O(1)$ windows are repaired on average, leading to $O(n)$ average time encoding and decoding. 

This paper is organized as follows. Section~\ref{sec:background} begins by providing background on periodicity, the previously-proposed codes, and the run-length-limited (RLL) constraint. Section~\ref{sec:codes} presents the proposed construction, and Section~\ref{sec:generalizations} explores several generalizations. Section~\ref{sec:combinatorical} provides a cardinality analysis, and finally Section~\ref{sec:conclusion} concludes this paper. \short{Some proofs are omitted and are available in the extended version~\cite{Extended}.}

% ---- Background ---- %
\section{Definitions, Preliminaries, and Related Works}
\label{sec:background}

We begin with several definitions and various results in the theory of periodicity, continue with the previous works for periodicity-constrained codes~\cite{chee2018coding, MultipleHeadRacetrack}, and conclude with background on the run-length-limited (RLL) constraint~\cite{MutuallyUncorrelated}.

% -- Notation -- %
\subsection{Notations}
\label{sec:background:notation}

For all $i$, we denote $[i] = \set{k\in\mathbb{N}}{1 \leq k \leq i}$. Let $\Sigma_q$ be a finite alphabet of size $q$ and let $\Sigma_q^n$ be the set of all vectors of length $n$ over $\Sigma_q$; without loss of generality, $0,1 \in \Sigma_q$. For a subset $A \subseteq \Sigma_q^n$, we denote by $\overline{A}$ the complement of $A$. For a vector $\v{s} = (s_1, \ldots , s_n)$ and $i, j \in [n], i \leq j$, we denote by $\v{s}_i^j$ the window $(s_i, \ldots , s_j)$ of $\v{s}$. A \emph{zero run} of length $r$ of a vector $\v{s} \in \Sigma_q^n$ is a window $\v{s}^{i+r-1}_i$, $i \in [n-r+ 1]$ such that $s_i = \cdots = s_{i+r-1} = 0$. The notation $\v{s}\v{t}$ denotes the concatenation of the two vectors
$\v{s}$ and $\v{t}$, and $\v{s}^k$ denotes the concatenation of $\v{s}$ with itself $k$ times. Unless stated otherwise, $\log$ refers to $\log_q$, where $q$ is the size of the alphabet. 

% -- Periodicity Definitions -- %
\subsection{Periodicity Definitions}
\label{sec:background:periodicityDefinitions}

We begin this subsection with a definition for the periodicity of a vector, continue by defining a periodicity-avoiding (PA) vector which avoids a specific period in all windows, and then extend this to a least-periodicity-avoiding (LPA) vector which avoids all periods up to a specific value in all windows.

\begin{definition} For $\v{s} \in \Sigma_q^n$, $p \in [n-1]$ is called a \emph{period} of $\v{s}$ if for all $i \in [n-p], s_{i} = s_{i+p}$.
\label{def:period}
\end{definition}

\begin{definition}[PA] For $\v{s} \in \Sigma_q^n$, $\v{s}$ is an \emph{$\ell$-window $p$-period avoiding vector} if every window of length $\ell$ does not possess period $p$. Let $B_q(n, \ell, p)$ be the set of such vectors, and let $b_q(n, \ell, p) = \abs{B_q(n, \ell, p)}$.
\label{def:PA}
\end{definition}

\begin{definition}[LPA] For $\v{s} \in \Sigma_q^n$, $\v{s}$ is an \emph{$\ell$-window $p$-least-period avoiding vector} if $\v{s}$ is an $\ell$-window $p'$-period avoiding (PA) vector for all $p' < p$. Equivalently, every window of length $\ell$ in $\v{s}$ does \emph{not} contain any period $p' < p$. Let $A_q(n, \ell, p)$ be the set of all such vectors $\v{s}$, and let $a_q(n, \ell, p) = \abs{A_q(n, \ell, p)}$. Notice that $A_q(n, \ell, p) = \bigcap_{p'=\lfloor (p+1)/2 \rfloor}^{p-1} B_q(n, \ell, p)$ as multiples of periods are periods. A code $\mathcal{C}$ is called an \emph{$(\ell,p)$-LPA code} if $\mathcal{C}\subseteq A_q(n, \ell, p)$. If the values of $\ell$ and $p$ are clear from the context, it is simply referred to as an \emph{LPA code}.
\label{def:LPA}
\end{definition}

\noindent This paper tackles the following three problems:

\begin{problem} Design $\ell$-window $p$-period LPA codes with efficient encoding/decoding that minimize the value of $\ell$ while requiring only a single redundancy symbol.
\label{prob:LPA}
\end{problem}
\begin{problem} Design $\ell$-window $p$-period LPA codes with efficient encoding/decoding that minimize the number of redundancy symbols for a given small value of $\ell$.
\label{prob:LPAred}
\end{problem}
\begin{problem} Study the values of $a_q(n, \ell, p)$ and $b_q(n, \ell, p)$.
\label{prob:card}
\end{problem}

% -- Periodicity Theorems -- %
\subsection{Theory of Periodicity}
\label{sec:background:periodicityTheorems}

Periodicity has been widely explored as a theoretical concept; we highlight key theorems utilized in Sections~\ref{sec:generalizations} and~\ref{sec:combinatorical}.

\begin{theorem}[Fine and Wilf's~\cite{FineWilf, choffrut1997combinatorics}]
Let $\v{s} \in \Sigma_q^n$ with periods $p_s$ and $p_t$ where $n \geq p_s + p_t - \gcd(p_s, p_t)$. Then $\gcd(p_s, p_t)$ is also a period of $\v{s}$.
\label{cor:gcd}
\end{theorem}

Theorem~\ref{cor:gcd} provides conditions for the uniqueness of a period in a message: if there are two periods $p_s,p_t < \lfloor n / 2 \rfloor + 2$, then $p_s$ and $p_t$ are both multiples of a smaller period ($\gcd(p_s, p_t)$). Therefore, by extending a message with a symbol that contradicts the minimal period, we find:

\begin{corollary}
Let $\v{s} \in \Sigma_q^n$. Then there exists $a \in \Sigma_q$ such that $\v{s}a \in \Sigma_q^{n+1}$ contains no periods less than $\lfloor n/2 \rfloor + 2$.\footnote{Notice that $n \geq 2p-4$ implies no periods less than $p$.}
\label{cor:primitive}
\end{corollary}

% -- Related Works -- %
\subsection{Related Works on Constrained Periodicity}
\label{sec:background:related}

Problem~\ref{prob:LPA} was first considered by Chee~\textit{et al.}~\cite{chee2018coding}, which presented a lower bound on $a_q(n, \ell, p)$ to prove that an LPA code with $\ell = \lceil \log(n) \rceil + p+1$ and a single redundancy symbol exists; yet, an explicit construction was not derived. Sima and Bruck~\cite{MultipleHeadRacetrack} later proposed an explicit construction with $O(n^2p\log n)$ time complexity for $\ell = \lceil \log(n)\rceil + 3p - 2$ using $p+1$ redundancy symbols; yet, the redundancy is significantly higher than Chee~\textit{et al.}~\cite{chee2018coding}. 
Section~\ref{sec:combinatorical} highlights the main results from Chee~\textit{et al.}~\cite{chee2018coding} regarding LPA codes, including the lower bound on $a_q(n,\ell,p)$ and a relationship between the PA constraint and the run-length-limited (RLL)~\cite{MutuallyUncorrelated} constraint.

\subsection{The Run-Length-Limited (RLL) Constraint}
\label{sec:background:RLL}

The \emph{run-length-limited} (\emph{RLL}) \emph{constraint} restricts the length of runs of consecutive zeros within encoded messages~\cite{MutuallyUncorrelated, marcus2001introduction}. Similar to~\cite{MutuallyUncorrelated}, we consider the \emph{$(0,k)$-RLL constraint}, which imposes the length of every run of zeros to be at most $k$, and for simplicity refer to this constraint as the \emph{$k$-RLL constraint}. Below is the definition of the constraint and the state-of-the-art construction for a single redundancy symbol.

\begin{definition}[RLL] A vector $\v{s} \in \Sigma_q^n$ satisfies the \emph{$k$-RLL constraint} if there are no zero runs of length $k$. Let $R_q(n, k)$ be the set of such vectors, and let $r_q(n, k) = \abs{R_q(n, k)}$. A code satisfying the $k$-RLL constraint is called a \emph{$k$-RLL code}.
\label{def:RLL}
\end{definition}

\begin{construction}[\hspace{-0.001ex}\cite{MutuallyUncorrelated}]
For all $n$ and $k = \lceil \log(n)\rceil + 1$, there exists an explicit construction of $k$-RLL codes with a single redundancy symbol and encoding/decoding with $O(n)$ time.
\label{const:RLL}
\end{construction} 

% ---- Proposed Scheme ---- %
\section{Single-Symbol LPA Construction}
\label{sec:codes}

This section tackles Problem~\ref{prob:LPA} through an approach of iteratively repairing invalid windows in the vectors, resulting in the following construction for a single redundancy symbol.
\begin{construction}
There exists an explicit construction of $(\ell,p)$-LPA codes for $\ell$ being the minimal value satisfying $\ell = \lceil \log(n-\ell+2)\rceil + p + 1$, a single redundancy symbol, and $O(n)$ average-time encoding and decoding complexity.
\label{const:core}
\end{construction}

The main idea is for the encoder to iteratively \emph{repair} invalid windows until no such windows exist, and then reverse these steps in the decoder. While this is relatively simple, the difficulty remains in proving its convergence due to the lack of monotonic progression: repairing a certain window may cause other previously-valid windows (both to the left and the right) to become invalid. Surprisingly, through a reduction to an acyclic graph walk, we nonetheless show that subtle properties of the repair routine inherently guarantee convergence.

This section continues by detailing the proposed encoder and decoder algorithms, proving their convergence through a reduction to an acyclic graph walk, and attaining $O(n)$ average time complexity. For the remainder of this section, $\ell$ is the minimal integer that satisfies $\ell = \lceil \log(n - \ell + 2)\rceil + p + 1$.

\subsection{Proposed Encoder and Decoder}
\label{sec:codes:encoderDecoder}

The encoder, which is explicitly described in Algorithms~\ref{alg:encoder} and~\ref{alg:routine}, iteratively removes invalid windows while appending a representation of the steps performed to the message. Inspired by Construction~\ref{const:RLL}, the redundancy symbol encodes whether any steps were taken: the symbol is initialized to one at the start, and becomes zero if a repair step is taken. The representation of a single step encodes the kernel of the periodic window removed (the first $p'$ symbols in a window with periodicity $p'$), the periodicity ($p'$), and the index of the window. Both the kernel and $p'$ are encoded within the same $p$ symbols by appending a one padded with zeros to the kernel. Notice that the \emph{message size is unchanged} as $\ell$ was chosen to satisfy $\ell = \lceil \log(n-\ell+2)\rceil + p + 1$. Overall, we proceed with such repair steps until there exists no invalid window. 

The decoder reverses the steps of the encoder, as inspired by the decoder from Construction~\ref{const:RLL}. The redundancy symbol is utilized to determine whether the last symbols of the message encode a step that was performed by the encoder. If so, then the decoder removes the step representation, reconstructs the invalid window by extending the given kernel according to the given period, and inserts it at the given index. 

Example~\ref{exam:mono} exemplifies the encoder for the binary case.
\begin{algorithm}[t]
    \centering
    \begin{algorithmic}[1]
    \renewcommand{\algorithmicrequire}{\textbf{Input:}}
    \renewcommand{\algorithmicensure}{\textbf{Output:}}
    \REQUIRE $\v{x} \in \Sigma_q^n$.
    \ENSURE $\v{y} \in \Sigma_q^{n+1}$ such that $\v{y} \in A_q(n+1, \ell, p)$. 
    
    \STATE $\v{y} \gets \v{x} 1$
    
    \WHILE{$\v{y} \notin A_q(n+1, \ell, p)$}
    
    \STATE $\v{y} \gets Repair(\v{y})$.
    
    \ENDWHILE
    
    \RETURN $\mathbf{y}$.
    
    \end{algorithmic}
    \caption{LPA Encoder}
    \label{alg:encoder}
\end{algorithm}

\begin{algorithm}[t]
    \centering
    \begin{algorithmic}[1]
    \renewcommand{\algorithmicrequire}{\textbf{Input:}}
    \renewcommand{\algorithmicensure}{\textbf{Output:}}
    \REQUIRE $\v{y} \in \Sigma_q^{n+1}$ such that $\v{y} \notin A_q(n+1, \ell, p)$.
    \ENSURE $\v{y} \in \Sigma_q^{n+1}$ such that $y_{n+1} = 0$.
    
    \STATE $i \gets $ index of first $\ell$-window in $\v{y}$ with period $p' < p$.
    
    \STATE Append $\v{y}_i^{i+p'-1}10^{p-p'-1}$ to the end of $\v{y}$ (i.e., $\v{y}_i^{i+p'-1}$, then one, then $p-p'-1$ zeros).
    
    \STATE Append the representation of $i$ (using $\lceil \log(n-\ell+2)\rceil$ symbols; zero-indexed) to $\v{y}$.
    
    \STATE Append $0$ to $\v{y}$.
    
    \STATE Remove the $\ell$-window at index $i$.
    
    \RETURN $\v{y}$.
    
    \end{algorithmic}
    \caption{$Repair$}
    \label{alg:routine}
\end{algorithm}
\begin{example}
Let $n=14$ and $p=4$ (thus $\ell = 8$) with
\begin{equation*}
    \v{x} = 10001010101100.
\end{equation*}
Algorithms~\ref{alg:encoder} and~\ref{alg:routine} perform the following steps:
\begin{enumerate}
    \item $\v{y} = \v{x}1 = 100010101011001$.
    \item $\v{y} \gets Repair(\v{y})$.
    \begin{enumerate}
        \item The 8-window starting at $i=3$ ($01010101$) is invalid as it possesses period $p'=2 < p$.
        \item $\v{y} = \v{y}0110^1 = 100010101011001\ 0110$.
        \item $\v{y} = \v{y}011 \hspace{9.5pt} = 100010101011001\ 0110\ 011$.
        \item $\v{y} = \v{y}0 \hspace{19.5pt} =  100010101011001\ 0110\ 011\ 0$.
        \item Remove the 8-window at index $i=3$ from $\v{y}$.
        \item Return $\v{y} = 100100101100110$.
    \end{enumerate}
    \item $\v{y} \gets Repair(\v{y})$.
    \begin{enumerate}
        \item The 8-window starting at $i=0$ ($10010010$) is invalid as it possesses period $p'=3 < p$.
        \item $\v{y} = \v{y}10010^0 = 100100101100110\ 1001$.
        \item $\v{y} = \v{y}000  \hspace{14.5pt} = 100100101100110\ 1001\ 000$.
        \item $\v{y} = \v{y}0  \hspace{24.5pt} = 100100101100110\ 1001\ 000\ 0$.
        \item Remove the 8-window at index $i=0$ from $\v{y}$.
        \item Return $\v{y} = 110011010010000$.
    \end{enumerate}
    \item Return $\v{y} = 110011010010000 \in A_2(15, 8, 4)$.
\end{enumerate}
\label{exam:mono}
\end{example}

Notice that the first call to the $Repair$ function in Example~\ref{exam:mono} created the invalid window which was then addressed by the second call. That is, while $Repair$ may fix the current window, it may also create other invalid windows. Therefore, it is unclear whether the algorithm will \emph{ever} converge, considering that each repair may lead to additional invalid windows. Indeed, we find that there even exist states ($\v{y} \in \Sigma_q^{n+1}$) that \emph{if ever reached} will cause Algorithm~\ref{alg:encoder} to never converge. This scenario is demonstrated in the following example.

\begin{example}
Let $n=14$ and $p=4$ (thus $\ell = 8$), with
\begin{equation*}
    \v{y} = 111111010101010.
\end{equation*}
The repair routine (Algorithm~\ref{alg:routine}) \emph{would} perform the following steps if $\v{y}$ is reached by Algorithm~\ref{alg:encoder} as an intermediate state:
\begin{enumerate}
    \item The window starting at $i=5$ ($10101010$) is invalid as it possesses period $p'=2 < p$.
    \item $\v{y} = \v{y}1010^1 = 111111010101010\ 1010$.
    \item $\v{y} = \v{y}101  \hspace{9.5pt} = 111111010101010\ 1010\ 101$.
    \item $\v{y} = \v{y}0 \hspace{19.5pt} = 111111010101010\ 1010\ 101\ 0$.
    \item Remove window at index $i=5$ from $\v{y}$.
    \item Return $\v{y} = 111111010101010$.
\end{enumerate}
That is, $Repair(\v{y}) = \v{y}$. Therefore, if Algorithm~\ref{alg:encoder} were to ever reach this $\v{y}$, then the encoder would never converge. 
\label{exam:loop}
\end{example}

Nonetheless, as proven in Section~\ref{sec:codes:convergence}, the proposed encoder always converges as it inherently avoids such intermediate states (e.g., Example~\ref{exam:loop}) due to subtle properties of the $Repair$ function. Further, Section~\ref{sec:codes:time} demonstrates that the number of steps taken is only $q-1=O(1)$ on average; thus, the encoder and decoder time complexity is $O(n)$ on average.

\subsection{Convergence Proof}
\label{sec:codes:convergence}

This section proves the convergence of the proposed encoder and decoder through a reduction to an acyclic graph walk. We show that the encoder inherently avoids intermediate states that will lead to infinite loops (e.g., Example~\ref{exam:loop}) by exploiting two subtle properties of the $Repair$ function: the fact that it is injective, and the fact that $Repair(\v{y})$ always ends with zero. The intuition for the proof is as follows. Let $\v{y}$ be given such that $Repair(\v{y}) = \v{y}$ (as in Example~\ref{exam:loop}), we show that the encoder will never reach such $\v{y}$ as an intermediate state. Since $Repair(\v{y}) = \v{y}$, then $\v{y}$ ends with zero; thus, the encoder will never start the repair steps with $\v{y}$. Further, \emph{as $Repair$ is injective}, then there exists no $\v{z}\neq \v{y}$ such that $Repair(\v{z}) = \v{y} = Repair(\v{y})$; thus, $\v{y}$ cannot be reached from a different intermediate state $\v{z}$. Therefore, the encoder will never reach any such $\v{y}$ as the encoder cannot start with such $\v{y}$ and the encoder will never update the intermediate state to be such $\v{y}$.

We generalize the above intuition in Theorem~\ref{the:encoderDefined} to also address cyclic structures that consist of more than one intermediate state (e.g., $Repair(\v{y}_1) = \v{y}_2$ and $Repair(\v{y}_2) = \v{y}_1$).

\begin{lemma} 
The $Repair$ function from Algorithm~\ref{alg:routine} is injective (that is, for all $\v{z} \neq \v{y}$, it holds that $Repair(\v{z}) \neq Repair(\v{y})$). 
\full{
\begin{IEEEproof} 
The inverse of $Repair$ on its image is given by decoding the kernel, $p'$ and $i$, and then reconstructing and inserting the window which was removed. As a unique window is defined by the kernel, $p'$, and $i$, then the inverse is well-defined. Therefore, the $Repair$ function is injective.
\end{IEEEproof}
}
\end{lemma}

\begin{theorem} 
The encoder from Algorithm~\ref{alg:encoder} is well-defined. 
\begin{IEEEproof}
Notice that if the encoder converges, then the output is in $A_q(n+1, \ell, p)$ by design, and thus a valid message is returned. The difficulty remains in proving the convergence. 

We model the encoder as a walk on a directed graph $G = (V,E)$ with nodes representing message states and edges representing the $Repair$ function. We let $S$ represent the possible start nodes of the algorithm. That is,
\begin{equation*}
V = \Sigma_q^{n+1} \quad\quad\quad S = \set{\v{v}1}{\v{v} \in \Sigma_q^n} \subseteq V,
\end{equation*}
\vspace{-5pt}
\begin{equation*}
E = \set{(\v{u}, \v{v})}{\v{u} \notin A_q(n+1,\ell,p),\; Repair(\v{u}) = \v{v}}.
\end{equation*}

Figure~\ref{fig:graph} illustrates an example of this graph. Notice that the in-degree of all nodes is at most one (as $Repair$ is injective), and that the in-degree of all nodes in $S$ is zero (as the output of $Repair$ always ends in 0). Assume by contradiction that there exists a cycle $C$ in $G$ that is reachable from a node in $S$. No node in $C$ belongs to $S$ as all nodes in $S$ have an in-degree of zero. Therefore, as $C$ is reachable from a node in $S$, then there exists an edge $\v{u} \rightarrow \v{v}$ such that $\v{u} \notin C$ and $\v{v} \in C$. Yet, this contradicts the in-degree of all nodes being at-most one (as there is another edge to $\v{v}$ from within the cycle).

\begin{figure}[t]
    \centering
    \includegraphics[width=0.965\linewidth]{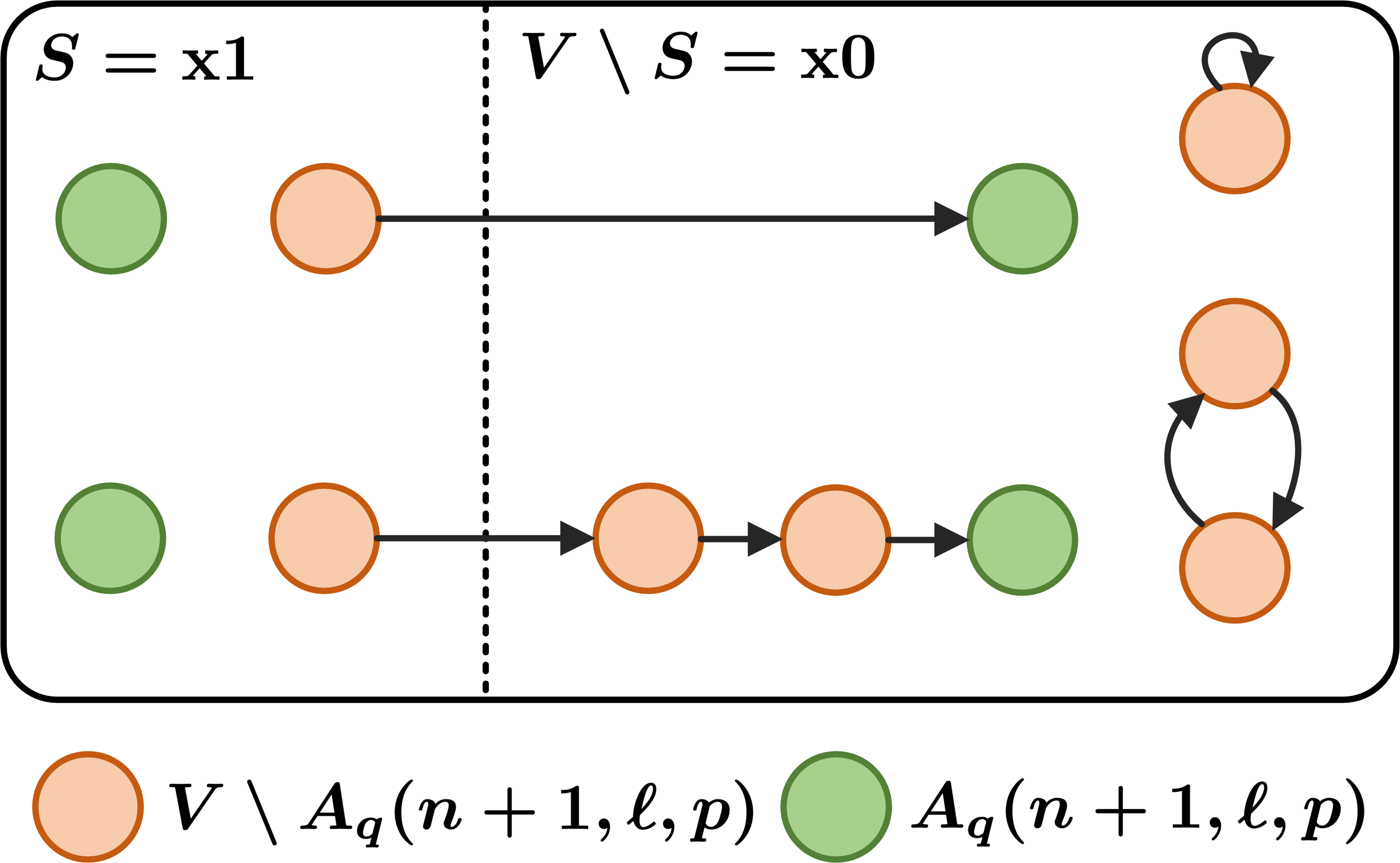}
    \caption{Example graph which Algorithm~\ref{alg:encoder} traverses. The algorithm starts in $S$ and applies the \emph{Repair} function until a valid node is reached (in $A_q(n+1,\ell,p)$). While cycles exist, they are unreachable from $S$ (see Theorem~\ref{the:encoderDefined}).}
    \label{fig:graph}
    \vspace{-10pt}
\end{figure}
Assume by contradiction that the encoder does not converge for some input $\v{x} \in \Sigma_q^{n}$. Let $\v{y}^{(1)}, \v{y}^{(2)},\ldots$ be the intermediate states of the encoder (the value of $\v{y}$ before each iteration of the while loop from Algorithm~\ref{alg:encoder}). Since $\v{y}^{(i)} \in \Sigma_q^{n+1}$ for all $i\in\mathbb{N}$ and $\abs{\Sigma_q^{n+1}} < \infty$, then there exist $i < j$ such that $\v{y}^{(i)} = \v{y}^{(j)}$. Therefore, by design of the encoder, we find that $\v{y}^{(i)} \rightarrow \v{y}^{(i+1)} \rightarrow \cdots \rightarrow \v{y}^{(j-1)} \rightarrow \v{y}^{(j)} = \v{y}^{(i)}$ is a cycle in $G$. We note that $\v{y}^{(i)}$ is reachable from a node in $S$ as $\v{y}^{(1)} \in S$ by properties of the encoder. Therefore, we found a cycle $C$ in $G$ that is reachable from a node in $S$. This is a contradiction.
\end{IEEEproof}
\label{the:encoderDefined}
\end{theorem}

\vspace{-5pt}

\begin{theorem}
The decoder is well-defined and correct.
\begin{IEEEproof}
The proof is similar to the proof of Construction~\ref{const:RLL}.
\end{IEEEproof}
\end{theorem}

\vspace{-5pt}

\subsection{Time Complexity}
\label{sec:codes:time}

This section extends the analysis of Section~\ref{sec:codes:convergence} to demonstrate that the average time complexity of the encoder and decoder is $O(n)$. We first show that the average number of steps is $O(1)$, and then propose an $O(n)$ algorithm for each step (i.e., the repair and inverse-repair functions). 

\begin{lemma}
The average number of iterations of the while loop in Algorithm~\ref{alg:encoder} is at most $q-1=O(1)$.
\begin{IEEEproof}
As shown in Theorem~\ref{the:encoderDefined}, an execution of Algorithm~\ref{alg:encoder} is equivalent to a walk on $G$. We notice that the two paths generated by two distinct inputs are disjoint as nodes in $G$ possess an in-degree of at most one. Thus, the sum of the lengths of all paths is the size of the union of all paths from all possible inputs. Therefore, as all paths are contained in $V\setminus S$ (excluding start nodes), the sum is bounded by $\abs{V\setminus S}$. Let $t(\v{x})$ be the length of the path for input $\v{x} \in \Sigma_q^n$; we find,
\begin{equation*}
    \sum_{\v{x} \in \Sigma_q^n} t(\v{x}) \leq \abs{V\setminus S} = q^{n+1} - q^n = (q-1) \cdot q^n.
\end{equation*}
Therefore, we find that the average path length is $q-1=O(1)$,
\begin{equation*}
    \frac{1}{q^n}\sum_{\v{x} \in \Sigma_q^n} t(\v{x}) \leq \frac{(q-1) \cdot q^n}{q^n} = q-1 = O(1).
\end{equation*}
\end{IEEEproof}
\label{lemma:O1}
\end{lemma}

\begin{corollary}
The encoder possesses $O(n)$ average time for $\ell \geq 2p-2$ and $O(np)$ average time otherwise.
\full{
\begin{IEEEproof}
We extend Lemma~\ref{lemma:O1} to prove the overall average time complexity of the encoder by proposing efficient worse-case algorithms for the window search in $Repair$. We propose two algorithms, corresponding to the cases of $\ell \geq 2p-2$ and $\ell < 2p-2$, with $O(n)$ and $O(np)$ time (respectively).

For $\ell \geq 2p-2$, we utilize the algorithm proposed by Main~\cite{main1989detecting} which decomposes the string using the $s$-factorization and then searches for the first occurrences of maximal periodicity\footnote{Maximal periodicity refers to periodic runs where extending the runs would contradict the periodicity. The relation to periodic $\ell$-windows is as follows: any periodic $\ell$-window is part of a maximal-periodicity run (can be extended to the left and right as much as possible), and any such run of length at least $\ell$ contains a periodic $\ell$-window in the first $\ell$ symbols (in particular).} ending within each factor. We modify the algorithm as follows to match the LPA constraint in this paper: the routine for step (3.1) is modified to only search for periods up to $p$ ($j$ iterates from $1$ to $p-1$ instead of $1$ to $n$) and windows of size at least $\ell$ (the condition $\text{LS}(j) + \text{LP}(j+1) \geq j$ is replaced with $\text{LS}(j) + \text{LP}(j+1) \geq \ell-j$), and the algorithm returns the index of the first $\ell$-window with periodicity.

For $\ell < 2p-2$, we exploit the equivalence provided by Chee~\textit{et al.}~\cite{chee2018coding} between the PA and RLL constraints through the difference function. That is, for each $p' < p$, we compute the $p'$-difference function, $d_{p'}: \Sigma_q^n \rightarrow \Sigma_q^{n-p'}$ where $(d_{p'}(\v{x}))_i = x_i - x_{i+p'}$, on the entire message $\v{y}$, and then check if this difference satisfies the $(\ell-p')$-RLL constraint (linear-time pass). This provides an $O(np)$ worst-case algorithm for the repair routine.
\end{IEEEproof}
}
\end{corollary}

\begin{corollary}
The decoder possesses $O(n)$ average time.
\full{
\begin{IEEEproof}
We notice that the inverse repair routine can be performed with $O(n)$ worst-case time complexity as all that is required is to decode $p'$, the kernel, and $i$, and then insert the reconstructed window of length $\ell = O(n)$. Therefore, as the number of iterations of the encoder is $O(1)$ (and the decoder performs the same number of iterations as the corresponding execution from the encoder), then we find that the time complexity of the decoder is $O(n)$ on average.
\end{IEEEproof}
}
\end{corollary}

\section{Extensions of the LPA Encoder}
\label{sec:generalizations}

This section tackles Problem~\ref{prob:LPAred} by proposing generalizations of Construction~\ref{const:core} to support smaller window sizes ($\ell < \lceil \log(n-\ell+2)\rceil + p + 1$) while minimizing the number of redundancy symbols. We demonstrate a trade-off between three proposed constructions which are all based on partitioning the input message into independent segments.

\begin{construction}
For any given $n, \ell, p$,
there exists an explicit construction for $(\ell,p)$-LPA codes with $k$ redundancy symbols, for $k$ the smallest integer such that $\ell \geq 2 \cdot (\lceil \log(n/k - \ell/2 + 2) \rceil + p + 1)$, and $O(n)$ average-time encoding/decoding.
\full{
\begin{IEEEproof}
Let $\v{x} \in \Sigma_q^n$ be the input message and let $\v{x}^{(1)}, \ldots, \v{x}^{(k)}$ be the partition of the vector into $k$ non-overlapping segments (e.g., $\v{x}^{(1)} = \v{x}_1^{n/k} = \v{x}_1 \cdots \v{x}_{n/k}$).\footnote{We assume without loss of generality that $k$ divides $n$. Otherwise, the last partition is of smaller size -- still attaining the desired LPA properties.} Let $\v{y}^{(1)}, \ldots, \v{y}^{(k)}$ be the encoded vectors for $\v{x}^{(1)}, \ldots, \v{x}^{(k)}$ according to Construction~\ref{const:core} with $\ell/2$ (respectively), and define $\v{y} = \v{y}^{(1)} \cdots \;\v{y}^{(k)}$. We now show that $\v{y} \in A_q(n+k, \ell, p)$.

Assume by contradiction that $\v{y} \notin A_q(n+k,\ell,p)$; thus, there exists an invalid window in $\v{y}$ of length $\ell$. As the invalid window is continuous, then at least $\ell/2 = \lceil \log(n/k-\ell/2+2) \rceil + p + 1$ symbols belong to the same segment $\v{x}^{(j)}$, for some $1\leq j \leq k$. Therefore, as a sub-vector of a periodic window is also periodic, we have found an invalid window of size $\lceil \log(n/k-\ell/2+2) \rceil + p + 1$ within the segment $\v{x}^{(j)}$. This is a contradiction to $\v{y}^{(1)}, \ldots, \v{y}^{(k)} \in A_q(n/k+1, \ell/2, p)$ by the correctness of Construction~\ref{const:core}. Therefore, $\v{y} \in A_q(n+k, \ell, p)$.

The time complexity of the proposed algorithm follows directly from that of Construction~\ref{const:core}.
\end{IEEEproof}
}
\label{const:2l}
\end{construction}
\begin{construction} 
For given $n, \ell, p$ such that $\ell \geq 3p-3$, there exists an explicit construction for $(\ell,p)$-LPA codes with $(p+3) \cdot (k-1)+1$ redundancy symbols, where $k$ is the smallest value that satisfies $\ell \geq \lceil \log(n/k-\ell+2) \rceil + p + 1$, and $O(n)$ average-time encoding/decoding.
\full{
\begin{IEEEproof}
Let $\v{x} \in \Sigma_q^n$ be the input message and let $\v{x}^{(1)}, \ldots, \v{x}^{(k)}$ be the partition of the message. Let $\v{y}^{(1)}, \ldots, \v{y}^{(k)}$ be the encoded vectors for $\v{x}^{(1)}, \ldots, \v{x}^{(k)}$ according to Construction~\ref{const:core} with $\ell$ (respectively). Define 
\begin{equation*}
\v{y} = \v{y}^{(1)}u^{(1)}\v{z}w^{(1)}\v{y}^{(2)}u^{(2)}\v{z}w^{(2)}\v{y}^{(3)}\cdots w^{(k-1)}\v{y}^{(k)}
\end{equation*}
where $\v{z} = 10\cdots 0 \in \Sigma_q^{p}$ (vector that does not possess any period) and, for all $j$, $u^{(j)}$ ($w^{(j)}$) are symbols chosen by Corollary~\ref{cor:primitive} to eliminate any periods less than $p$ from the last $2p-4$ symbols of $\v{y}^{(j)}$ (first $2p-4$ symbols of $\v{y}^{(j+1)}$). We now show that $\v{y} \in A_q(n+(p+3) \cdot (k-1)+1, \ell, p)$.

Assume by contradiction that $\v{y} \notin A_q(n+(p+3) \cdot (k-1)+1, \ell, p)$; thus, there exists an invalid window in $\v{y}$ of length $\ell$ with period $p' < p$. We divide into the following cases:
\begin{itemize}
    \item If the window is contained within one of $\v{y}^{(1)}, \ldots, \v{y}^{(k)}$. This is a contradiction to $\v{y}^{(1)}, \ldots, \v{y}^{(k)} \in A_q(n/k+1, \ell, p)$ by the correctness of Construction~\ref{const:core}.
    \item Else, if the window contains $\v{z}$. This is a contradiction as $\v{z}$ does not possess any period less than $p$, and thus the $i$-th window also cannot possess any period less than $p$.
    \item Else, we find that the window either contains the last $2p-4$ symbols of some $\v{y}^{(j)}$ with $u^{(j)}$, or $w^{(j)}$ with the first $2p-4$ symbols of some $\v{y}^{(j+1)}$ (as $\ell \geq 3p-3$, and the window does not contain $\v{z}$). This is a contradiction to the choice of $u^{(j)}, w^{(j)}$ using Corollary~\ref{cor:primitive}.
\end{itemize}
Therefore, $\v{y} \in A_q(n+(p+3) \cdot (k-1)+1, \ell, p)$.
The time complexity follows directly from that of Construction~\ref{const:core}.
\end{IEEEproof}
}
\label{const:p2}
\end{construction}

\begin{construction}
For given $n, \ell, p$ such that $\ell \geq 4p-7$, there exists an explicit construction for $(\ell,p)$-LPA codes with $3 \cdot k-2$ symbols of redundancy, where $k$ is the smallest value that satisfies $\ell = \lceil \log(n/k-\ell+2) \rceil + p + 1$, and $O(n)$ average-time encoding/decoding.
\full{
\begin{IEEEproof}
Let $\v{x} \in \Sigma_q^n$ be the input message and let $\v{x}^{(1)}, \ldots, \v{x}^{(k)}$ be the partition of the message. Let $\v{y}^{(1)}, \ldots, \v{y}^{(k)}$ be the encoded vectors for $\v{x}^{(1)}, \ldots, \v{x}^{(k)}$ according to Construction~\ref{const:core} with $\ell$ (respectively). Define 
\begin{equation*}
\v{y} = \v{y}^{(1)}u^1w^{1}\v{y}^{(2)}u^2w^{2}\v{y}^{(3)}\cdots w^{k-1}\v{y}^{(k)}
\end{equation*}
where, for all $j$, $u^j$ ($w^j$) are chosen by Corollary~\ref{cor:primitive} to eliminate any period less than $p$ from the last $2p-4$ symbols of $\v{y}^{(j)}$ (first $2p-4$ symbols of $\v{y}^{(j+1)}$). The proof that $\v{y} \in A_q(n+3\cdot k-2, \ell, p)$ is similar to that of Construction~\ref{const:p2}, where only the first and third cases are possible (as $\ell \geq 4p-7$). The time complexity follows directly from that of Construction~\ref{const:core}.
\end{IEEEproof}
}
\label{const:3}
\end{construction}

Overall, for given $n, \ell, p$, we seek the construction with minimal redundancy.  First, note that Construction~\ref{const:3} is preferable over Construction~\ref{const:p2} when $\ell \geq 4p-7$. Further, Construction~\ref{const:p2} is preferable over Construction~\ref{const:2l} when $\ell \geq 3p-3$ and
\begin{equation*}
    q^{\ell/2 - p - 1} + \frac{\ell}{2} - 2 \leq \frac{q^{\ell -p - 1} + \ell - 2}{p+3}.
\end{equation*}
Similarly, Construction~\ref{const:3} requires less redundancy than Construction~\ref{const:2l} when $\ell \geq 4p-7$ and
\begin{equation*}
    q^{\ell/2 - p - 1} + \frac{\ell}{2} - 2 \leq \frac{q^{\ell -p - 1} + \ell - 2}{3}.
\end{equation*}

% ---- Combinatorical ---- %
\section{Cardinally Analysis}
\label{sec:combinatorical}

This section analyzes the cardinality of the PA and LPA constraints, extending the analysis of Chee~\textit{et al.}~\cite{chee2018coding}. We begin with the first upper bound for $a_q(n,\ell,p)$ and a demonstration that $\ell=\log(n-2\ell+p)+p-3.5$ is a lower bound for single-symbol redundancy. We continue with several interesting exact formulas for the remaining cases not covered by the bounds.

% ---- Bounds ---- %
\subsection{Lower and Upper Bounds on \texorpdfstring{$a_q(n, \ell, p)$}{aq(n, l, p)}}
\label{sec:combinatorical:bounds}

We begin with results from~\cite{chee2018coding} in Theorems~\ref{the:cheeLower}~and~\ref{the:bRLL}, and then propose additional bounds via an RLL reduction.

\begin{theorem}[Chee~\textit{et al.}~\cite{chee2018coding}] For all $n, \ell, p$, and for all $q$,
\begin{equation*}
a_q(n, \ell, p) \geq q^n \cdot \left(1 - \frac{n}{(q-1)\cdot q^{\ell-p}}\right).
\end{equation*}
\label{the:cheeLower}
\end{theorem}
\vspace{-15pt}
In particular, for $\ell = \lceil \log(n) \rceil + p + 1$, we find $a_q(n, \ell, p) \geq q^{n-1}$ and thus a code with a single redundancy symbol exists.
\begin{theorem}[Chee~\textit{et al.}~\cite{chee2018coding}] For all $n, \ell, p$ and for all $q$,
\begin{equation*}
b_q(n, \ell, p) = q^p \cdot r_q (n-p, \ell-p)
\end{equation*}
\vspace{-15pt}
\label{the:bRLL}
\end{theorem}
\noindent We extend this result to the LPA constraint as follows,
\begin{lemma} For all $n, \ell, p$ and for all $q$,\footnote{Equality holds for $p \in \{2,3\}$ due to the result from Definition~\ref{def:LPA}.}
\begin{equation*}
a_q(n, \ell, p) \leq q^{p-1} \cdot r_q (n-p+1, \ell-p+1).
\end{equation*}
\begin{IEEEproof}
Via Theorem~\ref{the:bRLL} and $A_q(n,\ell,p) \subseteq B_q(n, \ell, p-1)$.
\end{IEEEproof}
\label{the:aRLL}
\end{lemma}
Therefore, by utilizing the bound on $k$-RLL codes in Theorem~\ref{the:RLL}, we find in Corollary~\ref{cor:aRLLSub} an upper-bound on $a_q(n,\ell,p)$.
\begin{theorem}[\hspace{-0.001ex}\cite{MutuallyUncorrelated}]
For all $n, k$ where $n\geq 2k$, and for all $q$,
\begin{equation*}
r_q(n,k) \leq q^{n-c \cdot \frac{n-2k}{q^k}}, \;\text{for}\; c = \frac{\log e(q-1)^2}{2q^2}.
\end{equation*}
\vspace{-15pt}
\label{the:RLL}
\end{theorem}
\begin{corollary} For all $n, \ell, p$, $n \geq 2\ell-p+1$, and for all $q$,
\begin{equation*}
a_q(n, \ell, p) \leq q^{n-c \cdot \frac{n-2\ell+p-1}{q^{\ell-p+1}}}, \;\text{for}\; c = \frac{\log e(q-1)^2}{2q^2}.
\end{equation*}
\vspace{-15pt}
\label{cor:aRLLSub}
\end{corollary}
We find the following corollary bounding the optimal window sizes for codes using a single redundancy symbol.
\begin{corollary}
For all $n, \ell, p$ where $n \geq 2\ell - p + 1$, and for all $q$, if there exists an $(\ell,p)$-LPA code with a single redundancy symbol, then $\ell \geq \log(n-2\ell+p) + p - 3.5$.
\full{
\begin{IEEEproof}
As there exists an $(\ell,p)$-LPA code with a single redundancy symbol, we conclude that 
\begin{equation*}
a_q(n + 1, \ell, p) \geq q^n.
\end{equation*}
We substitute the result from Corollary~\ref{cor:aRLLSub} to conclude that,
\begin{equation*}
q^{(n+1)- \frac{\log e(q-1)^2}{2q^2} \cdot \frac{(n+1)-2\ell+p-1}{q^{\ell-p+1}}} \geq q^n.
\end{equation*}
Hence,
\begin{align*}
& n + 1 - \frac{\log e(q-1)^2}{2q^2} \cdot \frac{(n+1)-2\ell+p-1}{q^{\ell-p+1}} \geq n \\
\Longleftrightarrow\;\; & \frac{\log e(q-1)^2}{2q^2} \cdot \frac{(n+1)-2\ell+p-1}{q^{\ell-p+1}} \leq 1 \\ 
\Longleftrightarrow\;\; & \log\left(\frac{\log e(q-1)^2}{2q^2} \cdot \frac{(n+1)-2\ell+p-1}{q^{\ell-p+1}}\right) \leq 0 \\
% \implies & \log\left(\frac{\log e (q-1)^2}{2q^2}\right) + \log((n+1)-2\ell+p-1) - \ell + p - 1 \leq 0 \\ 
\Longleftrightarrow\;\; & \ell \geq \log(n-2\ell+p) + p - 3 + \log\left(\frac{\log e(q-1)^2}{2}\right).
\end{align*}
We notice that $\log\left(({\log e(q-1)^2})/{2}\right) \geq -1/2$ for $q \geq 2$, and thus $\ell \geq \log(n-2\ell+p) + p - 3.5$.
\end{IEEEproof}
}
\label{cor:coreLPA}
\end{corollary}
Therefore, we find that Construction~\ref{const:core} is near the lower bound of the optimal construction. In particular, if $n \geq 3\ell-2p+2$, then we differ by up to 5.5 from the lower bound.
% \[ \log(n-2\ell+p) - \log(n-\ell+2) = \log\left(\frac{n-2\ell+p}{n-\ell+2}\right) \]
% \[ = \log\left(1 - \frac{\ell-p+2}{n-\ell+2}\right) \]
% \[ \frac{\ell-p+2}{n-\ell+2} \leq \frac{1}{2} \iff 2\ell-2p+4 \leq n-\ell+2 \]
% \[ \iff 3\ell-2p+2 \leq n \]

% ---- Exact ---- %
\subsection{Exact Formulas}
\label{sec:combinatorical:exact}
Here, we provide interesting exact formulas for the cases of $n=\ell$ and $n \leq 2\ell-2p+4$. We begin with $b_q(n,n,p)$.
\begin{lemma}
For all $n, p$, and for all $q$,
\begin{equation*}
b_q(n, n, p) = q^n - q^p.
\end{equation*}
\short{\newpage}
\full{
\begin{IEEEproof}
We show $|{\overline{B_q(n,n,p)}}| = q^p$, from which the desired expression follows. We notice that all vectors in $\overline{B_q(n,n,p)}$ are defined exclusively by their first $p$ symbols (as the vector contains periodicity $p$), and that any choice of $p$ symbols for the beginning of a vector can be extended to length $n$. Therefore, there are exactly $q^p$ vectors in $\overline{B_q(n,n,p)}$.
\end{IEEEproof}
}
\label{the:bNN2}
\end{lemma}
We now address the more challenging case of $a_q(n, n, p)$.
\begin{theorem}
For all $n, p$ such that $n \geq 2p-4$, and for all $q$, 
\begin{equation*}
a_q(n, n, p) = q^n - \frac{q}{q-1} \cdot \sum_{d=1}^{p-1} \mu(d) \cdot \left(q^{\left\lfloor \frac{p-1}{d}\right\rfloor} - 1\right).
\end{equation*}
where $\mu$ is the M\"{o}bius function.
\full{
\begin{IEEEproof}
Recall from Definition~\ref{def:LPA} that $A_q(n,n,p) = \bigcap_{p'=1}^{p-1} B_q(n,n,p')$; thus, we find by inclusion-exclusion,
\begin{small}
\begin{multline*}
\abs{\overline{A_q(n,n,p)}} =  \sum_{k=1}^{p-1} (-1)^{k+1} \cdot \left(\sum_{\substack{S \subseteq [p-1]\\ \abs{S} = k}}\abs{\bigcap_{j\in S}{\overline{B(n, j)}}}\right).
\end{multline*}
\end{small}
By Theorem~\ref{cor:gcd}, we note that $\bigcap_{j\in S}{\overline{B(n, j)}} \subseteq \overline{B(n, \gcd (S))}$. Further, $\overline{B(n, \gcd (S))} \subseteq \bigcap_{j\in S}{\overline{B(n, j)}}$ follows trivially as multiples of a period are also periods. Therefore,
\begin{multline*}
\abs{\overline{A_q(n,n,p)}} =  \sum_{k=1}^{p-1} (-1)^{k+1} \cdot \left(\sum_{\substack{S \subseteq [p-1]\\ \abs{S} = k}}\abs{{\overline{B(n, \gcd (S))}}}\right).
\end{multline*}
As the inner summation is only dependent on $\gcd(S)$, then we present this equivalent summation,
\begin{multline*}
\abs{\overline{A_q(n,n,p)}} = \sum_{g=1}^{p-1} \abs{\overline{B(n,n,g)}} \\   \cdot \left[\sum_{k=1}^{p-1}(-1)^{k+1}
\abs{\set{S \subseteq [p-1]}{\substack{\abs{S}=k,\\ \gcd(S) = g}}}\right].
\end{multline*}
Through properties of $\gcd$ and the results of Nathanson~\cite{nathanson2007affine},
\begin{multline*}
\abs{\overline{A_q(n,n,p)}} = \sum_{g=1}^{p-1} \abs{\overline{B(n,n,g)}} \\   \cdot \left[\sum_{d=1}^{\left\lfloor \frac{p-1}{g} \right\rfloor} \mu(d) \sum_{k=1}^{\left\lfloor\left\lfloor \frac{p-1}{g} \right\rfloor/d\right\rfloor}(-1)^{k+1} \binom{\left\lfloor \left\lfloor \frac{p-1}{g} \right\rfloor/d \right\rfloor}{k}\right].
\end{multline*}
We note that the inner-most summation is equal to $1$; thus,
\begin{equation*}
\abs{\overline{A_q(n,n,p)}} = \sum_{g=1}^{p-1} \abs{\overline{B(n,n,g)}} \cdot \left[\sum_{d=1}^{\left\lfloor \frac{p-1}{g} \right\rfloor} \mu(d) \right].
\end{equation*}
Substituting the result from Lemma~\ref{the:bNN2} and rearranging the summation leads to the desired result,
\begin{equation*}
\abs{\overline{A_q(n,n,p)}} = \frac{q}{q-1}\cdot \sum_{d=1}^{p-1} \mu(d) \cdot (q^{\lfloor(p-1)/d\rfloor} - 1).
\end{equation*}
\end{IEEEproof}
}
\label{the:aNN}
\end{theorem}

This result can be extended for more cases when $n > \ell$.
\begin{theorem} For all $n, \ell, p$ such that $n \leq 2\ell - 2p+4$,
\begin{equation*}
\abs{\overline{A_q(n,\ell,p)}} = \abs{\overline{A_q(\ell, \ell, p)}} \cdot q^{n-\ell} \cdot (1 + (n-\ell) \cdot (1- q^{-1})).
\end{equation*}
% \begin{equation*}
% a_q(n,\ell,p) = q^{n-\ell} a_q(\ell,\ell,p) - (n-\ell) \cdot q^{n-\ell-1} \cdot (q-1)\cdot \abs{\overline{A_q(\ell,\ell,p)}} .
% \end{equation*}
\full{
\begin{IEEEproof}
Let $n, \ell, p$ be given, and let $i = n - \ell$. We first propose a decomposition of $A_q(\ell, \ell, p)$, and we then utilize that to determine the cardinality of $A_q(n, \ell, p)$. We decompose $A_q(\ell,\ell,p)$ according to the length of the shortest suffix in the vector that avoids all periodicity up to $p$:
\begin{equation*}
A_q(\ell,\ell,p) = \bigcup_{j=\ell-i}^{\ell} S_j,
\end{equation*}
while $S_{\ell-i}$ denotes the set of all vectors in $A_q(\ell,\ell,p)$ where the last $\ell-i$ symbols belong to $A_q(\ell-i, \ell-i, p)$, and $S_j$ for $j > \ell - i$ is the set of all vectors such that the last $j-1$ symbols belong to $\overline{A_q(j-1, j-1, p)}$, yet the last $j$ symbols belong to $A_q(j, j, p)$. The union is disjoint as if the last $j$ symbols belong to $A_q(j, j, p)$, then the last $j'\geq j$ also belong to $A_q(j', j', p)$. Notice that $\abs{S_{\ell-i}} = a_q(\ell-i, \ell-i, p)\cdot q^i$ as any vector in $a_q(\ell-i, \ell-i, p)$ can be extended with any $i$ symbols, and $\abs{S_j} = \abs{\overline{A_q(j-1, j-1, p)}} \cdot (q-1) \cdot q^{\ell-j} = \abs{\overline{A_q(\ell, \ell, p)}} \cdot (q-1) \cdot q^{\ell-j}$ for $j > \ell-i$ by Corollary~\ref{cor:primitive} (as only a single symbol continues the periodicity) and as $\overline{A_q(k, k, p)}$ is independent of $k$ when $k \geq 2p-4$ (see Theorem~\ref{the:aNN}). 

%\newpage
We now consider $a_q(n, \ell, p)$ by utilizing the decomposition for the first $\ell$ symbols of vectors in $A_q(n, \ell, p)$. Specifically, we find two cases: (1) when the first $\ell$ symbols belong to $S_{\ell-i}$, then the remaining $i$ symbols may be chosen freely, (2) when the first $\ell$ symbols belong to $S_{j}$ for $j > \ell-i$, then the next $\ell-j+1$ symbols are chosen to contradict the period (which is defined by the last $2p-4$ symbols of the first $\ell$ symbols), and the remaining symbols are chosen freely. That is, we find
\begin{equation*}
a_q(n, \ell, p) = \abs{S_{\ell-i}} \cdot q^{i} + \sum_{j=\ell-i+1}^{\ell} \abs{S_j} \cdot (q^{\ell-j+1}-1) \cdot q^{j-(\ell-i+1)}
\end{equation*}
\vspace{-10pt}
\begin{multline*}
= a_q(\ell-i, \ell-i, p)\cdot q^{2i} \; + \\ \abs{\overline{A_q(\ell, \ell, p)}}\cdot q^{i-1}\cdot  \underbrace{\sum_{j=\ell-i+1}^{\ell} (q-1) \cdot (q^{\ell-j+1}-1)}_{=q(q^i-1) - iq + i}.
\end{multline*}
The desired result follows from the above expression.
% \begin{multline*}
% = a_q(\ell-i, \ell-i, p)\cdot q^{2i} \; + \\ \abs{\overline{A_q(\ell, \ell, p)}}\cdot q^{i-1}\cdot  \left(q(q^i-1) - iq + i\right)
% \end{multline*}
% \begin{multline*}
% = a_q(\ell-i, \ell-i, p)\cdot q^{2i} \; + \\ \abs{\overline{A_q(\ell, \ell, p)}}\cdot \left(q^i(q^i-1) - iq^i + iq^{i-1}\right)
% \end{multline*}
% \begin{multline*}
% = q^{\ell+i} + \abs{\overline{A_q(\ell, \ell, p)}}\cdot \left(-q^i - iq^i + iq^{i-1}\right)
% \end{multline*}
\end{IEEEproof}
}
\label{the:aNN+1}
\end{theorem}

\short{\vspace{-5pt}}

\section{Conclusion}
\label{sec:conclusion}
In this work, we study codes that constrain periodicity within windows of the encoded messages. We propose a construction with a single symbol of redundancy based on iteratively repairing invalid windows until a valid message is encountered. Even though the algorithm does not possess monotonic progression, we prove convergence with linear average time complexity through a reduction to an acyclic graph walk. We continue by generalizing the proposed construction to offer a trade-off between the window size and the number of additional redundancy symbols. Lastly, we study the cardinality of the constraints to both prove that the proposed construction is nearly optimal, and to mention novel exact formulas. Overall, we establish foundational constructions for constrained periodicity that may be fundamental for many different applications, such as racetrack memories.

\section*{Acknowledgments}
The research was Funded by the European Union. Views and opinions expressed are however those of the authors only and do not necessarily reflect those of the European Union or European Research Council.

\bibliographystyle{IEEEtran}
\bibliography{refs}

\begin{thebibliography}{10}
\providecommand{\url}[1]{#1}
\csname url@samestyle\endcsname
\providecommand{\newblock}{\relax}
\providecommand{\bibinfo}[2]{#2}
\providecommand{\BIBentrySTDinterwordspacing}{\spaceskip=0pt\relax}
\providecommand{\BIBentryALTinterwordstretchfactor}{4}
\providecommand{\BIBentryALTinterwordspacing}{\spaceskip=\fontdimen2\font plus
\BIBentryALTinterwordstretchfactor\fontdimen3\font minus
  \fontdimen4\font\relax}
\providecommand{\BIBforeignlanguage}[2]{{%
\expandafter\ifx\csname l@#1\endcsname\relax
\typeout{** WARNING: IEEEtran.bst: No hyphenation pattern has been}%
\typeout{** loaded for the language `#1'. Using the pattern for}%
\typeout{** the default language instead.}%
\else
\language=\csname l@#1\endcsname
\fi
#2}}
\providecommand{\BIBdecl}{\relax}
\BIBdecl

\bibitem{Racetrack2008}
S.~S.~P. Parkin, M.~Hayashi, and L.~Thomas, ``Magnetic domain-wall racetrack
  memory,'' \emph{Science}, 2008.

\bibitem{RacetrackPIEEE}
R.~Bläsing \emph{et~al.}, ``Magnetic racetrack memory: From physics to the
  cusp of applications within a decade,'' \emph{Proc. of the IEEE}, 2020.

\bibitem{HiFi}
C.~Zhang \emph{et~al.}, ``{Hi-Fi} playback: Tolerating position errors in shift
  operations of racetrack memory,'' \emph{SIGARCH Comput. Archit. News}, 2015.

\bibitem{chee2018coding}
Y.~M. Chee \emph{et~al.}, ``Coding for racetrack memories,'' \emph{IEEE TIT},
  2018.

\bibitem{CheeReconstruction}
------, ``Reconstruction from deletions in racetrack memories,'' in \emph{IEEE
  ITW}, 2018.

\bibitem{MultipleHeadRacetrack}
J.~Sima and J.~Bruck, ``Correcting deletions in multiple-heads racetrack
  memories,'' in \emph{IEEE ISIT}, 2019.

\bibitem{MutuallyUncorrelated}
M.~Levy and E.~Yaakobi, ``Mutually uncorrelated codes for {DNA} storage,''
  \emph{IEEE TIT}, 2019.

\bibitem{FineWilf}
N.~J. Fine and H.~S. Wilf, ``Uniqueness theorems for periodic functions,''
  \emph{Proceedings of the American Mathematical Society}, 1965.

\bibitem{choffrut1997combinatorics}
C.~Choffrut and J.~Karhum{\"a}ki, ``Combinatorics of words,'' in \emph{Handbook
  of Formal Languages}.\hskip 1em plus 0.5em minus 0.4em\relax Springer, 1997,
  pp. 329--438.

\bibitem{marcus2001introduction}
B.~H. Marcus, R.~M. Roth, and P.~H. Siegel, ``An introduction to coding for
  constrained systems,'' \emph{Lecture notes}, 2001.

\bibitem{main1989detecting}
M.~G. Main, ``Detecting leftmost maximal periodicities,'' \emph{Discrete
  Applied Mathematics}, vol.~25, no. 1-2, pp. 145--153, 1989.

\bibitem{nathanson2007affine}
M.~B. Nathanson, ``Affine invariants, relatively prime sets, and a phi function
  for subsets of $\{$1, 2,..., n$\}$,'' \emph{Integers}, vol.~7, p.~A1, 2007.

\end{thebibliography}
\end{document}